\documentclass[floats,aps,prb,draft,preprint,showpacs]{revtex4}
\usepackage{epsf}
\begin{document}
%\tighten
%\draft
%\wideabs{
\title{Stability of Driven Josephson Vortex Lattice in Layered Superconductors Revisited}
\author{S.N.Artemenko and S.V.Remizov}
%\address{
\affiliation{ Institute for Radioengineering and Electronics of
Russian Academy of Sciences, Mokhovaya 11, 103907 Moscow, Russia }

\date{\today}

%}

\begin{abstract}
We analytically study stability of sliding lattice of Josephson
vortices driven by a transport current in the stack direction in
strong in-plane magnetic field. In contrast to recent findings we
obtain that there are no diverse configurations of stable vortex
lattices, and, hence, the stable sliding vortex lattice can not be selected by
boundary conditions. We find that only the triangular (rhombic)
lattice can be stable, its stability being limited by a critical
velocity value. At higher velocities there are no simple stable
lattices with single flux line per unit cell. Oblique sliding
lattices are found to be never stable. Instability of such
lattices is revealed beyond the linear approximation in
perturbations of the lattice.
\end{abstract}

\maketitle

\section{Introduction}

Strong anisotropy of layered superconductors, such as high-T$_c$
superconductors, results in a number of specific features. In
particular, magnetic field parallel to the conducting layers
induces coreless vortices similar to fluxons in conventional
Josephson tunnel junctions~\cite{Bulaevskii1, BulaevskiiClem}. In
contrast to the fluxons in conventional Josephson junctions, the
current and magnetic field of a Josephson vortex in layered
superconductors are spread over many conducting layers. The
transport current flowing in the stack direction forces the
vortices to slide along superconducting layers. Since in contrast
to standard Abrikosov vortices the Josephson vortices are
coreless, their motion does not involve perturbations of the
amplitude of the superconducting order parameter and corresponds
at low temperatures to an underdamped regime. So the velocity of such
vortices, in principle, can be quite large. Many dynamical
properties of Josephson vortices in layered superconducting
crystals are similar to those of fluxons in artificially prepared
stacked Josephson junctions (see, \textit{e. g.},
Ref.~\onlinecite{Kr}). The Josephson vortices can be arranged in a
lattice, however, possibility of multiple metastable states
corresponding to vortex configurations which do not form any
lattice is predicted as well at small magnetic field~\cite{Le}.
At larger magnetic field the Josephson vortices form a lattice
and, hence, in the flux-flow regime the coherent motion of the
vortex lattice is expected. Such a regime must induce coherent
electromagnetic radiation from the uniformly sliding lattice. The
flux-flow regime of this lattice was observed in BSCCO mesa
structures~\cite{LeeNorman,LeeGuptasarma,HechtfisherKleiner1,HechtfisherKleiner2,Latyshev}.
However, the regime of coherent sliding of the lattice was found
to be limited by a maximum voltage and, hence, by a maximum
lattice velocity~\cite{HechtfisherKleiner1,HechtfisherKleiner2,LeeGuptasarma,Latyshev}.
Above this voltage either a different regime was observed~\cite{HechtfisherKleiner1, HechtfisherKleiner2} or no stable I-V
curves were found in a close range of currents and voltages~\cite{LeeGuptasarma}. In addition, above the maximum velocity a
broad-band non-Josephson emission in the microwave region was
observed~\cite{HechtfisherKleiner1, HechtfisherKleiner2}.

The upper limit for the velocity of the coherent flux-flow regime
was obtained theoretically in our studies of stability of the
vortex lattice~\cite{AR-Fr,AR-Jap}, and was related to the interaction
of the vortices with Josephson plasma modes excited by moving
vortices. A regular motion of Josephson vortices in a form of the
rhombic lattice, usually called the triangular lattice, was found
to be stable at velocities up to a critical value $v_{cr}
\simeq c s/(2\lambda_\| \sqrt{\epsilon})$, where $s$ is the period of the
crystal lattice in the stack direction, $\lambda_\|$ is the London
penetration depth, $\epsilon$
is the dielectric constant in the stack direction. This critical
velocity was identified with the experimentally observed velocity
limiting two regimes of the vortex
motion~\cite{HechtfisherKleiner1,HechtfisherKleiner2,LeeGuptasarma,Latyshev}.
We have found that at higher velocities sliding of a regular vortex
lattice is frustrated due to a growth of fluctuations of the vortex
lattice. The instability is induced by interaction of the vortices
with the Josephson plasma mode. The similar results were obtained
both in the limit of large magnetic fields, when the distance
between the vortices is smaller than the size of the non-linear
region near the vortex center, and for the case of lower magnetic
fields~\cite{AR-Fr,AR-Jap} when the vortices are separated by
distances much larger than the size of the non-linear region.

However, qualitatively different conclusions on the stability of
moving vortex lattice were made in recent papers by Koshelev and
Aranson~\cite{KoshelevAranson1, KoshelevAranson2}. According to
their calculations a stable vortex lattice motion can occur not
only in a form of the triangular lattice, but also in a form of
various oblique lattices (see fig.~\ref{figLattice}), a particular
experimentally realized lattice being uniquely selected by
boundary conditions. Regimes of stable motion of various oblique
vortex lattices were found also for velocities exceeding the
critical value $v_{cr}$. The existence of a set of various stable
vortex lattices for a given sliding velocity would lead to
interesting physical consequences. Since a particular vortex
lattice is uniquely selected by boundary conditions, the radiation
spectrum must also strongly depend on boundary conditions.
Specifically, in contrast to other vortex lattices, the triangular
one does not emit at the main radiation frequency related to the
time needed to shift the lattice by its period. This happens
since the emissions by adjacent vortex rows, which in the
triangular lattice are shifted by the half-period, compensate each
other. Therefore, in case of the triangular lattice the radiation
is generated at even harmonics only. The oblique lattices, in
which the shifts between the adjacent vortex rows are not equal to
the half-period, can radiate both at even and odd harmonics as
well. Furthermore, the intensity of the radiation maximum
predicted at the Josephson plasma frequency~\cite{AR-Fr,AR-Jap}
would be much larger if there exists a stable vortex lattice
arrangement at $v > v_{cr}$.

To eliminate the discrepancy and to resolve the problem we
reconsider stability of sliding vortex lattices and find that the
oblique vortex lattices are not stable. Instability of oblique
lattices is found out from the analysis of equations of motion for
perturbations of the lattice beyond the linear approximation in
the perturbations (Sec. IV). We illustrate this statement by a
particular case of the equilibrium state, \textit{i. e.}, of the
lattice at rest (Sec. III). In the equilibrium state the results
are especially transparent, because the stable vortex arrangement
can be selected as that corresponding to a minimal energy. We
conclude that oblique lattices correspond to a saddle point of the
energy functional with respect to perturbations of a uniform
lattice, and only the triangular lattice corresponds to a minimum.
This is revealed when the terms beyond the quadratic ones are kept
in the corresponding energy expansion. Therefore, to study
stability of sliding vortex lattices by means of equations of
motion one should proceed beyond the linear approximation with
respect to perturbations.

\section{Main equations}

Electrodynamic properties of layered superconductors can be
described in terms of gauge invariant potentials which can be
treated as superconducting momentum in layer $n$, ${\bf p}_n$, and
gauge invariant phase difference between layers $n+1$ and $n$ of
the superconducting order parameter, $\varphi_n$. In general case
electric and magnetic fields in the superconductor depend also on
gauge-invariant scalar potential $\mu_n=(1/2)\partial_t \chi_n +
\Phi_n$, $\Phi_n$ is the electric potential in the $n$-th layer.
We do not take into account here the components of the electric
field related to $\mu_n$ since we consider low temperatures when
the effects of branch imbalance related to $\mu_n$ can be ignored.

Equations for the moving vortex lattice are derived similar to our
previous paper~\cite{AR-JL}, namely, substituting the expression
for the current density ${\bf j}$ presented as a sum of
superconducting and quasiparticle currents to the Maxwell equation
$$
  \nabla \times {\bf H}
  =
  \frac{4 \pi}{c}{\bf j}
  +
  \frac{1}{c}
  \partial_t {\bf D}.
$$
Then we find equations for ${\bf p}_n$ and $\varphi_n$ which have
a form:
\begin{eqnarray}
  &&
     \frac{c}{2 s}\partial_{xx}^2\varphi_n -
  c \frac{\partial_{x}p_{n+1}-\partial_{x}p_n}{s} =
  \frac{4 \pi}{c}
  \left[
    j_c \sin \varphi_n +
    \frac{\sigma_{\perp}}{2 s} \partial_t \varphi_n
  \right] +
  \frac{\epsilon}{2 c s }\partial_{tt}^2\varphi_n  \label{eqMainSystemOrig1}
  \\
  &&
   - \frac{c}{2 s^2}(\partial_{x}\varphi_{n} - \partial_{x}\varphi_{n-1}) +
  \frac{c}{s^2}\left(p_{n+1} + p_{n-1} - 2p_n\right) =
  \frac{4 \pi}{c}
  \left[
    \frac{c^2}{4\pi\lambda_{\|}^2} p_n
    + \sigma_{\|}\partial_t p_n
  \right]
,  \label{eqMainSystemOrig2}
\end{eqnarray}
Here $\lambda_\|$ and $\lambda_\perp$ are screening length for
screening currents flowing in the plane and stack directions,
respectively, $\sigma_{\|}$ and $\sigma_{\perp}$ are quasiparticle
conductivities for related directions.

Excluding superconducting momenta from
Eqs.~(\ref{eqMainSystemOrig1}-\ref{eqMainSystemOrig2}) one readily
obtains the equation of motion for superconducting phases
$\varphi_n$
\begin{equation}
  \left[
     \frac{s^2}{\lambda_\|^2} \left(
      1 + \frac{\Omega_r}{\Omega_p^2} \partial_t \right) - \partial_n^2
  \right]
  \left(
    \partial_{tt}^2 \varphi_n +
    \omega_r \partial_t \varphi_n +
    \omega_p^2 \sin \varphi_n
  \right)
   =
  \frac{s^2}{\lambda_\|^2}
  \left(
     1 + \frac{\Omega_r}{\Omega_p^2} \partial_t
  \right)
    \partial_{xx}^2 \varphi_n
.\label{eqMainPhi}
\end{equation}
Here $\partial_n^2 f_n = f_{n+1} + f_{n-1} - 2
f_n$ corresponds to the discrete version of the second derivative.
$\Omega_r = 4 \pi i \sigma_\|$, $\omega_r =  4 \pi \sigma_\perp /
\epsilon$ are relaxation frequencies. $\Omega_p$ and $\omega_p$
are plasma frequencies for the directions along and perpendicular
to the layers, $\Omega_p = c/\lambda_\|$ is much larger than all
typical frequencies of the considered problem, $\omega_p =
c/\sqrt{\epsilon}\lambda_\perp \ll \Omega_p$.

Studies of the vortex dynamics both in our papers and in papers by
Koshelev and Aranson are based upon such equation.

\begin{figure}[!ht]
  \vskip 0mm
  \epsfxsize=70mm% \epsfysize=50mm
  \centerline{\epsffile{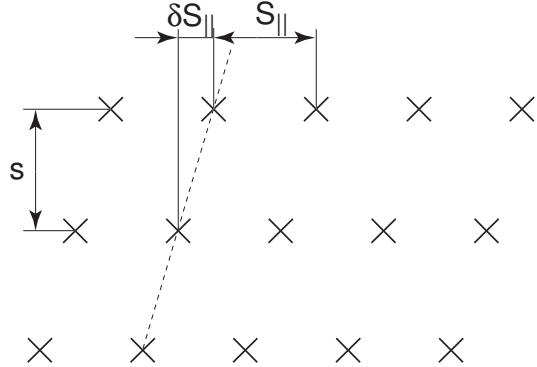}}
  \caption{
The configuration of the vortex lattice is defined by $\chi = 2
\pi \delta S_\|/S_\|$. The rectangular lattice corresponds to
$\chi = 0$, the oblique lattices are defined by $0 < \chi < \pi$,
and the triangular lattice corresponds to $\chi = \pi$.
  }
\label{figLattice}
\end{figure}

We limit our study by the case of large magnetic field considered
in Refs.~\cite{KoshelevAranson1, KoshelevAranson2}. In high
magnetic field the cores of Josephson vortices, \textit{i. e.},
non-linear regions at the centers of Josephson vortices, strongly
overlap and solution of Eq.~(\ref{eqMainPhi}) can be found
perturbatively as a sum of a linear term and a small oscillating
correction:
\begin{equation}
  \varphi_n^{(0)}=
  Y_n +
  \psi_n(Y_n)
.\label{eqDefPsi}
\end{equation}
Here $Y_n = q_0 x - \omega_0 t + \chi n$, $\omega_0 = q_0 v$, $q_0
= 2 \pi / S_\|$, $S_\|$ is the period of the chain of vortices
along the superconducting layers, and $\chi$ specifies the type of
the vortex lattice (see fig.~\ref{figLattice}).

\begin{equation}
  \psi_n(Y)
  =
  \mathop{\Re} {\cal G}(\omega_0, q_0, \chi) e^{i Y};
  \qquad
  {\cal G}
  = 
  \frac{i \omega_p^2 K^2(\omega_0, \chi)}{\hat c^2(\omega_0) q_0^2 - K^2(\omega_0, \chi) (\omega_0^2 + i \omega_0 \omega_r)}
.\label{psi}
\end{equation}
Here $K(\omega, q_\perp)^2 = 1 - i \Omega_r \omega/ \omega_p^2 + 4
(\lambda_\|/s)^2 \sin^2 q_\perp/2$, $\hat c^2(\omega) = c^2
(1 - i \Omega_r \omega / \omega_p^2)/\sqrt{\epsilon}$. 
The denominator of ${\cal G}(\omega_0,
q_0, \chi)$ does not become zero since $\omega_0 = q_{0} v$
and vortex velocity is small enough if below the critical value $v_{cr}$
mentioned above and discussed below in Sec. IV. Hence, for stable 
vortex lattice $|\psi_n(Y)| \ll 1$. 
%Since $\omega_0 = q_{0} v$
%if vortex velocity is small enough denominator of ${\cal G}(\omega_0,
%q_0, \chi)$ does not become zero and, hence, $|\psi_n(Y)| \ll
%1$. This condition is satisfied for stable vortex lattice, {\it
%i.e.}, for sliding velocities below the critical value $v_{cr}$
%mentioned above and discussed below in Sec. IV.

\section{Stationary vortex lattice}

At first we consider the vortex lattice at rest as the most simple
and physically transparent case. Stability of a stationary vortex
lattice is determined by its free energy. Minimum of the free energy
as function of $\chi$ corresponds to the stable vortex lattice.
The energy of the vortex lattice can be considered as a
sum of the energy of magnetic field induced by superconducting
electrons,
$$
  E^{(H)} =  \sum\limits_n \int dx H^2/8 \pi
,
$$
of the kinetic energy of superconducting currents
$$
  E^{(s)} =  \sum\limits_n \int dx \frac {m}{2 n_s e^2} j_s^2
,
$$
and the  Josephson energy of $n$ Junctions.
%\cite{xxxShmidt}.
$$
  E^{(J)} = \sum\limits_n \frac{ \Phi_0 j_c c}{2 \pi}(1 - \cos\varphi_n)
.
$$
Then the total energy can be readily presented as
\begin{equation}
   E^{(tot)}= \sum\limits_n \int dx \left[
    \frac{4 c^2}{8 \pi s} \left(
      \frac{(\varphi_n')^2}{4} - \varphi_n' (p_{n+1} - p_n) + (p_{n+1} - p_n)^2 + \frac{p_n^2}{l^2}
    \right)
          +
       \frac{j_c}{2} (1 - \cos \varphi_n)
  \right]
.
  \label{eqEnergy}
\end{equation}

The variation of the energy functional (\ref{eqEnergy}) over $p_n$
and $\varphi_n$ yields
Eqs.~(\ref{eqMainSystemOrig1}-\ref{eqMainSystemOrig2}) with
dissipative terms equal to zero.

\begin{figure}[!t]
  \vskip 0mm
  \epsfxsize=75mm
  \centerline{\epsffile{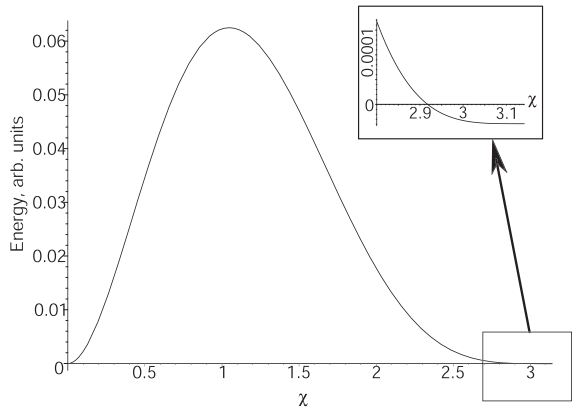}}
  \caption{
Dependence of the vortex energy on $\chi$. The curve is calculated
for $\lambda_\| / s = 160$.
  }
\label{figStabEn}
\end{figure}

To determine the stable vortex configuration we substitute
expressions for $p_n(Y_n)$ and $\varphi_n^{(0)}$
(\ref{eqDefPsi}-\ref{psi}) to equation for the energy density
(\ref{eqEnergy}). After some algebra the expression for the energy
density of the vortex lattice reads
\begin{equation}
 E^{(tot)}= \frac{ c^2}{8 \pi s} \left(
   \frac{2 l^4 \sin^2(\frac{\chi}{2})^2 (\sin(\frac{\chi}{2}) - 1)^2}{\lambda_{\perp}^4 q_0^2}
   %\right.
   -   \frac{l^2 \sin^2(\frac{\chi}{2})}{\lambda_{\perp}^4 q_0^2}
   +
    %\left.
   \displaystyle
   \frac{1}{4} q_0^2
   -
   \frac{1}{8 \lambda_{\perp}^4 q_0^2}
   +
   \frac{1}{2 \lambda_{\perp}^2}
 \right)
. \label{eqEnergyFin}
\end{equation}
From Eq.~(\ref{eqEnergyFin}) one can see that the energy minimum
corresponds to $\chi=\pi$. It means that in accordance with
previous results (see, \textit{e. g.},
Ref.~\onlinecite{BulaevskiiClem}), only the triangular lattice is
stable (fig.~\ref{figStabEn}). The difference between the energies
of the triangular and the rectangular lattices is small and proportional
to $q_0^{-2}$.

\section{Stability of sliding Josephson vortex lattice}

Now we consider a sliding vortex lattice. We adopt here the method
used in Ref.~\onlinecite{VolkovGlen} to study a system of two
coupled Josephson junctions. The solution of (\ref{eqMainPhi}) for
$\varphi_n$ can be presented in a form
\begin{equation}
  \varphi_n(x,t)=\varphi_n^{(0)}(Y)+\delta\varphi_n(x,t)
,\label{eqHiMain}
\end{equation}
where $\varphi_n^{(0)}(Y)$ given by
Eqs.~(\ref{eqDefPsi}-\ref{psi}) describes the uniform motion of
the vortex lattice with the velocity $v$, and $\delta
\varphi_n(x,t)$ corresponds small perturbations of the uniformly
sliding lattice, $|\delta\varphi_n(x,t)| \ll 1$.

The vortex lattice is stable if all possible perturbations $\delta
\varphi_n (x,t)$ decrease with time and there are no increasing
solutions for $\delta \varphi_n (x,t)$. If in the linear
approximation in $\delta \varphi_n (x,t)$ there are solution which
neither decay nor increase, then the linear approximation in
perturbation is not sufficient and one should take into account
higher order perturbations in $\delta \varphi_n (x,t)$.

The equation for $\delta\varphi$ in the linear approximation we
derive substituting Eq.~(\ref{psi}) into Eq.~(\ref{eqMainPhi}) and
performing Fourier transformation over $x,\,t$ and $n$. It reads
\begin{eqnarray}
   &&F(\omega, q_{\|}, q_{\perp})
   \delta\varphi(\omega, q_{\|}, q_{\perp})
      +
      f(\omega + \omega_0, q_{\|} + q_0, q_{\perp} + \chi)
   \delta\varphi(\omega + \omega_0, q_{\|} + q_0, q_{\perp} + \chi)
   +\nonumber
   \\
   &&
   f^* (\omega - \omega_0, q_{\|} - q_0, q_{\perp} - \chi)
   \delta\varphi(\omega - \omega_0, q_{\|} - q_0, q_{\perp} - \chi) = 0
. \label{eqSpecLin1}
\end{eqnarray}
Here $q_\|$ and $q_\perp$ ($|q_\perp| < \pi$) are the wave vectors
in the directions parallel and perpendicular to the superconducting
layers, respectively, $F(\omega, q_{\|}, q_{\perp}) = \hat
c(\omega)^2 q_{\|}^2 - K^2(\omega, q_{\perp})(\omega^2 + i \omega \omega_r
- i \omega_p^2 {\mathop{\Im\,}} {\cal G}(\omega_0, q_0, \chi))$, $f(\omega, q_{\|}, q_{\perp}) =
K^2(\omega, q_{\perp}) \omega_p^2 / 2$.

The values of the perturbation with shifted arguments,
$\delta\varphi(\omega \pm \omega_0, q_{\|} \pm q_0, q_{\perp} \pm \chi)$, are small compared to
$\delta\varphi(\omega , q_{\|}, q_{\perp})$ with small values of
the arguments. So using perturbation approach with respect to
$S_\| / \lambda_J$ and neglecting $\delta\varphi(\omega \pm 2 \omega_0,
q_\| \pm 2 q_0, q_\perp \pm 2 \chi)$, we express the
shifted values in terms of $\delta\varphi(\omega, q_{\|}, q_{\perp})$ and
find a simple equation of motion for $\delta\varphi$ which determines the
eigenmodes of the vortex lattice.
\begin{eqnarray}
  &&
   \left( F(\omega, q_{\|}, q_{\perp}) -
    \frac
    {
      f^*(\omega, q_{\|}, q_{\perp})
      f(\omega + \omega_0, q_{\|} + q_0, q_{\perp} + \chi)
    }
    {F(\omega + \omega_0, q_{\|} + q_0, q_{\perp} + \chi)}
  \right. \nonumber\\
  &&
    \left. - \frac
    {
      f(\omega, q_{\|}, q_{\perp})
      f^*(\omega - \omega_0, q_{\|} - q_0, q_{\perp} - \chi)
    }
    {F(\omega - \omega_0, q_{\|} - q_0, q_{\perp} - \chi)
    }
  \right)\delta\varphi(\omega, q_{\|}, q_{\perp})
  = 0
. \label{eqSpecLin}
\end{eqnarray}
In this expression two last terms originate from interaction of
the first harmonic of oscillating field induced by the sliding
lattice with the Josephson plasma mode.

As it follows from Eq.(\ref{psi}) ${\cal G}(\omega_0, q_0, \chi)
\propto (S_\|/\lambda_J)^2$ is small, and one may neglect terms
with ${\cal G}(\omega_0, q_0, \chi)$ in the expressions for $F$ in
the denominators of the last terms of Eq.(\ref{eqSpecLin}). Such
simplifications allow us to calculate from Eq.(\ref{eqSpecLin})
the dispersion relation defining the spectrum of collective
oscillations
\begin{eqnarray}
 && \omega^2
  =
  -i\omega \omega_r
  +
  \frac{\hat c^2(\omega)}{K^2(\omega, q_\perp)} +
    \omega_p^2  \nonumber \\
 && \times \mathop{\Im} \left(
     i {\cal G}(\omega_0, q_0, \chi)
    -
    \frac{i}{2} {\cal G}(\omega + \omega_0, q_\| + q_0, q_\perp +
    \chi)
    -
    \frac{i}{2} {\cal G}(\omega - \omega_0, q_\| - q_0, q_\perp -
    \chi)  \right).
\label{eqM1}
\end{eqnarray}

Consider first the triangular vortex lattice, $\chi = \pi$. In the
long-wavelength limit, for the lattice in rest we obtain the
sound-like spectrum with the velocities $c_\|= c/\sqrt{\epsilon}$ and
$c_\perp= (\lambda S_\| / \sqrt{8\epsilon}\pi \lambda_\perp^2) c$
in the in-plane and in stack directions, respectively. These
velocities coincide with ones obtained by Volkov~\cite{Volkov}.
For the sliding triangular lattice the imaginary part of the
frequency of the eigenmodes, which determines the decrement of
damping, is positive for small velocities $v < v_{cr} \simeq
(s/2\sqrt{\epsilon}\lambda_\|) c
 % \left(\sqrt{1+R/4}- \sqrt{R/4} \right)
.$ Thus at such velocities the triangular lattice is stable.

As it follows from Eq.~(\ref{eqM1}) at larger velocities for any
value of $\chi$ there are values of $q_\perp$ for which the
imaginary part of the frequency of the oscillations becomes
positive. This means that fluctuations increase with time
resulting in an instability of the lattice at such velocities. The
value of $v_{cr}$ is about $3 \times 10^5 \mbox{\it m/c}$ for
typical parameters $\lambda_\| / s = 100$ and $\epsilon \approx
20$. This estimate of the critical velocity coincides with the
velocity limiting flux-flow regime in Refs.~\onlinecite{HechtfisherKleiner1,LeeGuptasarma}.

Now we consider in more details the case of arbitrary $\chi$ and
$v<v_{cr}$. The sign of the imaginary part of $\omega$, which
determines the decrement of damping, sets conditions for stability
of the vortex lattice. A positive damping factor for a given
$\chi$ and for all values of $q_{\|}$ and $q_{\perp}$ means the
stability of the vortex lattice. While a negative damping factor,
in other words, a positive increment of growth, manifests the
instability. Analytical and numerical calculations show that the
damping factor is not negative for the whole region of values of
$\chi$ for a given value of the vortex velocity including zero velocity.
This result was interpreted in Refs.~\onlinecite{KoshelevAranson1,
KoshelevAranson2} as an evidence of the stability of these
lattices. However, the existence of a set of different stable lattices
at zero velocity apparently contradicts to the studies of the
equilibrium lattice~\cite{BulaevskiiClem} and to the
results of the preceding section according to which only the
triangular lattice should be stable. The point is that though for
the oblique lattices the damping factor is positive for finite
values of ${\bf q}$, it is equal to zero for ${\bf q} = 0$. This
indicates that the linear approximation on $\delta \varphi$ does
not give an ultimate answer on the stability problem, and further study is
needed. The situation can be clarified considering the particular
example of the equilibrium state when the stable solution can be
selected using the energy considerations. If the series expansion
of the energy with respect to perturbations near the extremum does
not contain the quadratic term, then contribution of the potential
energy to the linearized equation of motion becomes zero
suggesting a zero damping coefficient in a corresponding equation
of motion. Therefore in order to study stability in such a case
one must take into account the next order perturbations in the
energy expansion in perturbations. If the third order term in
perturbations in the expansion of the potential energy is non-zero
then the extremum of the energy corresponds to a saddle point and,
hence, to an instability.

These considerations cannot be directly applied to the
nonequilibrium case the stability of which should be studied by
means of equations of motion, but they point out that to find
whether solutions with zero damping factor in the linear
approximation are really stable, one must include in the equation
of motion for the phase the next order terms in perturbations
$\delta\varphi$. Keeping in Eq.(\ref{eqMainSystemOrig2}) linear
and quadratic terms in perturbations we find the equation for
$\delta\varphi$ which has the form
\begin{eqnarray}
  &&
   \left[
    \hat c(\omega)^2 q_{\|}^2 - K^2(\omega, q_{\perp})(\omega^2 + i \omega \omega_r - i
    \omega_p^2 \mathop{\Im} \left({\cal G}(\omega, q_{\|}, q_{\perp})\right))
  \right] \delta\varphi_{(0)} +
  \nonumber \\
  &&
   \frac{\omega_p^2}{2}
  \left[
%    K^2(\omega +\omega_0, q_{\perp} + \chi)\delta\varphi_{(+1)} +
    K^2_{(+1)}\delta\varphi_{(+1)} +
%  \right.
%  \nonumber \\
%  &&
%    \left.
%    K^2(\omega - \omega_0, q_{\perp} - \chi)\delta\varphi_{(-1)}
    K^2_{(-1)}\delta\varphi_{(-1)}
  \right] +
%  \nonumber \\
%  &&
  \omega_p^2 \mathop{\Re}\left({\cal G}(\omega, q_{\|}, q_{\perp})\right)
    K^2(\omega, q_{\perp})\left[
      \delta\varphi^2_{(-0)}
    \right] +
  \nonumber \\
  &&
  \frac{\omega_p^2}{2 i}
  \left[
%    K^2(\omega + \omega_0, q_{\perp} + \chi)\left[
    K^2_{(+1)}\left[
      \delta\varphi^2_{(+1)}
    \right] -
%  \right.
%  \nonumber \\
%  &&
%  \left.
%    K^2(\omega - \omega_0, q_{\perp} - \chi)\left[
    K^2_{(-1)}\left[
      \delta\varphi^2_{(-1)}
    \right]
  \right]
  = 0
. \label{eqSpecNLin}
\end{eqnarray}
Here $\delta\varphi_{(0)} = \delta\varphi(\omega, q_{\|},
q_{\perp})$, $\delta\varphi_{(\pm 1)} = \delta\varphi(\omega \pm
\omega_0, q_{\|} \pm q_0, q_{\perp} \pm \chi)$,
$K^2_{(\pm 1)} = K^2(\omega \pm \omega_0, q_{\perp} \pm \chi)$,
  and
$\left[\delta\varphi^2_{(\pm1)}\right]$ denote the Fourier
transforms of $[\delta\varphi_n(t,x)]^2$ with arguments $(\omega
\pm \omega_0, q_{\|} \pm q_0, q_{\perp} \pm \chi)$.

Eq.~(\ref{eqSpecNLin}) can be studied by perturbation approach in
$\delta\varphi$ and $(S_\|/\lambda_J)^2$ similar to previous study
of Eq.~(\ref{eqSpecLin1}). Furthermore, we limit our analysis of
the nonlinear case by the study of stability with respect to the
uniform perturbations, since the stability of the state with
$\textbf{q} = 0$ only was not ascertain in the linear
approximation in $\delta\varphi$. As the first approximation for
$\delta\varphi$ we choose a solution corresponding to the
perturbation of the parameter $\chi$ describing the tilt of the
lattice:
$$
  \delta\varphi^{(1)}
  =
  \frac{1}{2} \mathop{\Re}(
    [\delta\chi +
      \delta(
        {\cal G}(\omega_0, q_0, \chi)
      )
    ] e^{i Y}
  )
  =
  \frac{\delta\chi}{2}  \mathop{\Re}((1 + {\cal G}_1 (\chi) \sin \chi) e^{i Y}).
$$
Here ${\cal G}_1 (\chi) = \partial {\cal G}(\omega_0, q_0, \chi)/ \partial (\sin \chi/2)$. Terms with $e^{2
i Y}$ and $e^{- 2 i Y}$ give small corrections to
$(\delta\varphi^{(1)})^2$ and should be neglected. Finally, the
equation for vortex lattice perturbation (\ref{eqSpecNLin}) in the
second order can be rewritten as
$$
  (\omega^2 + i \omega \omega_r) \delta\varphi^{(2)} = 2 \mathop{\Re} {\cal G}_1(\chi) \sin (\chi) (\delta\chi)^2
.
$$
Since the right-hand size of this equation does not depend on
$\omega$ and $\mathop{\Re} {\cal G}_1(\chi) \ne 0$ for any value of $\chi$, there
are always increasing perturbations $\delta\varphi^{(2)}$ unless
the factor before $(\delta\chi)^2$ is equal to zero, \textit{i.
e.}, when $\chi$ is a multiple of $\pi$. As it was found above in
the linear approximation in lattice perturbations the rectangular
lattice, $\chi=0$, is not stable due to perturbations with finite
$q_\perp$, so there is only one simple vortex lattice with one
flux quantum per cell which could be stable, namely, the
triangular one, $\chi=\pi$.

\section{Conclusions}

In the frame of a simple theoretical model we studied stability of sliding
lattices of Josephson vortices induced by strong magnetic field applied
parallel to conducting layers of layered superconductor and driven by a
transport current in the stack direction. We found that in contrast to
recent theoretical predictions~\cite{KoshelevAranson1, KoshelevAranson2}
the type of the stable sliding lattice is not selected by boundary
conditions, but there is only one possible stable vortex lattice
arrangement of sliding Josephson vortices containing single flux line per
unit cell, namely, the triangular lattice. In other words, an
experimentalist cannot have an influence on the structure of the sliding
vortex lattice manipulating with experimental conditions. Sliding of
vortices in the form of the regular lattice is found to be stable up to the
critical velocity only. At higher velocities there is no sliding regime
with the vortices coherently arranged in an oblique, triangular or
rectangular vortex lattices.

\section{Acknowledgments}

This work was supported by Russian Foundation for Basic Research
(projects 01-02-17527 and  02-02-06322) and by Russian state
program on superconductivity.

\end{document}